\documentclass[aps,prb,superscriptaddress,twocolumn,showpacs,longbibliography]{revtex4-1}
\usepackage[utf8]{inputenc}
\usepackage{amsmath}
\usepackage{amsthm, amssymb} 
\usepackage{bbold}                       
\usepackage{graphicx}
\usepackage{color}
\usepackage{times, verbatim}  
\usepackage{braket}
\usepackage{bm}
\usepackage[caption=false]{subfig}

\usepackage{hyperref}
\hypersetup{colorlinks=true}

\AtBeginDocument{%
    \newwrite\bibnotes
    \def\bibnotesext{Notes.bib}
    \immediate\openout\bibnotes=\jobname\bibnotesext
    \immediate\write\bibnotes{@CONTROL{REVTEX41Control}}
    \immediate\write\bibnotes{@CONTROL{%
    apsrev41Control,author="08",editor="1",pages="0",title="0",year="1"}}
     \if@filesw
     \immediate\write\@auxout{\string\citation{apsrev41Control}}%
    \fi
}%

\newcommand{\BEq}{\begin{eqnarray}}
\newcommand{\EEq}{\end{eqnarray}}

\begin{document}

\title{Two-particle Correlation Functions in Cluster Perturbation Theory: Hubbard Spin Susceptibilities}

\affiliation{Department of Physics, Virginia Tech, Blacksburg, Virginia 24061, USA}
\affiliation{Computational Sciences and Engineering Division and Center for Nanophase Materials Sciences, Oak Ridge, Tennessee 37831, USA}

\author{P.T. Raum$^1$}
\author{G. Alvarez$^2$}
\author{Thomas Maier$^2$}
\author{V. W. Scarola$^1$}

\begin{abstract}
Cluster Perturbation Theory (CPT) is a computationally economic method commonly used to estimate the momentum and energy resolved single-particle Green’s function.  It has been used extensively in direct comparisons with experiments that effectively measure the single-particle Green’s function, e.g., angle-resolved photoemission spectroscopy. However, many experimental observables are given by two-particle correlation functions.  CPT can be extended to compute two-particle correlation functions by approximately solving the Bethe-Salpeter equation.  We implement this method and focus on the transverse spin-susceptibility, measurable via inelastic neutron scattering or with optical probes of atomic gases in optical lattices. We benchmark the method with the one-dimensional Fermi-Hubbard model by comparing with known results.
\end{abstract}

\date{\today}

\maketitle

\section{Introduction}

Cluster methods offer viable approximations to otherwise intractable  quantum many-body problems\cite{Maier2005}.  Strongly interacting models pose challenging problems at first glance because the Hilbert space scales exponentially.  Yet certain problems are tractable because small sub-systems offer good approximations to the thermodynamic limit.  Such problems present an opportunity to approximate eigenstates as factorized subsystems.  In lattice models the subsystems can be real-space clusters.  

Cluster methods are often applied to Hubbard models with local (on-site) Hubbard interactions:
\begin{align}
H_{\text{H}} = H_0+U\sum_{i}n_{i,\uparrow}n_{i,\downarrow}
\label{eq_hubbard_gen}
\end{align}
where $H_0$ is a single-particle term that includes inter-site hopping. 
$U$ is the Hubbard interaction parameter and $n_{i,\sigma}=c^{\dagger}_{i,\sigma}c_{i,\sigma}$, where $c^{\dagger}_{i,\sigma}$ creates a fermion at site $i$ and spin state $\sigma\in(\uparrow,\downarrow)$.  This model is written in a single band but cluster methods can be adapted to include multiband models as well.  In the following we focus on a single band and nearest neighbor hopping of energy $t$.

CPT \cite{Senechal2000,Senechal2002a,Gros1993,Zacher2002,Senechal2012} is one of the simplest cluster methods applicable to $H_{\text{H}} $. It uses exact diagonalization results from small clusters and perturbatively couples them together using inter-cluster perturbation theory, thus offering approximations to the thermodynamic limit.  The CPT formalism allows a straightforward estimate of important single-particle correlation functions, e.g., the spectral function.  The spectral function can be measured in angle resolved photo emission spectroscopy thus offering a useful connection between Eq.~\ref{eq_hubbard_gen} and experiments on, for example, transition metal oxides \cite{Damascelli2003}.

Two-particle correlation functions, by contrast, offer important insight into certain ordered phases of $H_{\text{H}} $.  Density-density and spin-spin correlations signal charge density and spin ordering, respectively.  Spin susceptibilities in particular are observable using neutron scattering in materials \cite{Lovesey1984,Dai2015}.  Interestingly, ultra-cold atoms in the Hubbard regime have been probed using optical analogues of neutron scattering to reveal spin-spin correlation functions in a Hubbard system as well \cite{Hart2015}.  It is therefore important that we have a straightforward method to approximate two-particle correlation functions.  

Previous work examined promotion of CPT to two-particle correlation functions. Kung et al. \cite{Kung2017} calculate the spin-susceptibility for the translational symmetry broken cluster and periodize the result but neglect inter-cluster terms. Brehm et al. \cite{Brehm2010} use the variational cluster approximation, an improvement over CPT, to approximately solve the Bethe-Salpeter equation for a two-dimensional (2D) Hubbard model  at large interaction strength, $U=8t$.  The variational cluster approximation includes a self consistent loop that improves accuracy but comes with the added cost of computational complexity.  Here we follow their approach but for CPT to construct a method that can be used with low computational cost so that it can be applied to more complex models, e.g., multi-band models.  

Following Ref.~\onlinecite{Brehm2010}, we extend the CPT formalism to two-particle correlation functions by using the Bethe-Salpeter equations to couple cluster solutions to the full lattice in direct analogy to the CPT method used for one-particle correlation functions.  This generalization allows us to compute spin-susceptibilities on small clusters that can in turn be used to compare with experiments done in the thermodynamic limit.   
Our central aim is to address issues in solving the Bethe-Salpeter equation by benchmarking the two-particle CPT method against known results in the single-band 1D Hubbard model.  The 1D Hubbard model has strong quantum fluctuations yet many exact (or nearly exact) results are known from Bethe Ansatz \cite{Lieb2003,1385i}, density matrix renormalization group (DMRG) \cite{White1992,White93}, and perturbative limits \cite{Muller1981,Bougourzi1996}.  At half filling we find that the spin-susceptibilities computed using CPT offer excellent approximations to known results for weak and strongly interacting limits.  We also find reasonable agreement at intermediate interaction strengths.  Away from half filling, we find more disagreement between CPT and DMRG as expected from strong charge fluctuations.  We note that care must be taken to apply the method to models in perturbative limits where CPT itself is accurate.  Our results set the stage for use of CPT to efficiently approximate two-particle correlation functions in perturbative limits of higher dimensional Hubbard models and models with band degrees of freedom where CPT is accurate.   

 The paper is organized as follows:  In Sec.~\ref{sec_onepart} we review one-particle CPT to establish notation and prepare for the generalization to two-particle correlation functions.  In Sec.~\ref{sec_twopart} we use the one-particle CPT formalism to define the protocol for two-particle CPT \cite{Brehm2010,Kung2017}.  In Sec.~\ref{sec_1dhubbard} we test the accuracy and viability of the formalism with the 1D Hubbard model in a single band at half filling by comparison with the Random Phase Approximation (RPA), DMRG calculations, and the $\text{M{\"u}ller}$ estimate \cite{Muller1981}.  In Sec.~\ref{sec_awayhalf} we compare CPT with DMRG away from half filling in the 1D Hubbard model.  We summarize in Sec.~\ref{sec_summary}.

\section{One-Particle Cluster Perturbation Theory}
\label{sec_onepart}

We first review the basics of one-particle CPT to establish notation and provide a framework for a direct application to two-particle correlation functions.  One-particle CPT focuses on approximating the single particle Green's function:

$$ G_{\alpha\beta}(i\nu) = \int_0^{1/k_BT}d\tau e^{i\nu\tau}\langle T_{\tau}c_\alpha(\tau)c_\beta^\dagger(0)\rangle$$
where $c_\beta^\dagger(\tau)$ $(c_\alpha(\tau))$ creates (annihilates) fermions in states indexed by $\beta$ ($\alpha$) at imaginary time $\tau$.  These composite indices specify both position, $\bm{r}_i$, and spin, $\sigma$, e.g., $\alpha \equiv (i,\sigma)$.  $T_{\tau}$ indicates time ordering.  Here we choose to present the formalism in the Matsubara representation at non-zero temperature $T$ for simplicity but when we apply the method numerically in Sec.~\ref{sec_1dhubbard}, we will pass to zero temperature.

The CPT scheme approximates the single-particle Green's function by breaking up the original lattice and calculating a Green's function in a mixed representation.  The mixed representation is given by tiling the lattice into clusters connected perturbatively by inter-cluster single-particle coupling in $H_0$.  The scheme results in a single-particle Green's function with a continuously valued momentum derived from a small system.  This is achieved formally by rewriting original lattice position (momentum) vectors as the sum of the cluster and superlattice position (momentum) vectors as: 
$ {\bf r}_i = {\bf R} + {\bf r}_a $  ($ {\bf k}  = {\bf \tilde{k}} + {\bf K }$), where $a$ indexes cluster sites and $\tilde{\bf k}$ superlattice momentum. This procedure breaks translational symmetry on the cluster but keeps it among clusters.

\begin{figure}[t]
	\subfloat{
\includegraphics[width=8cm,angle=0]{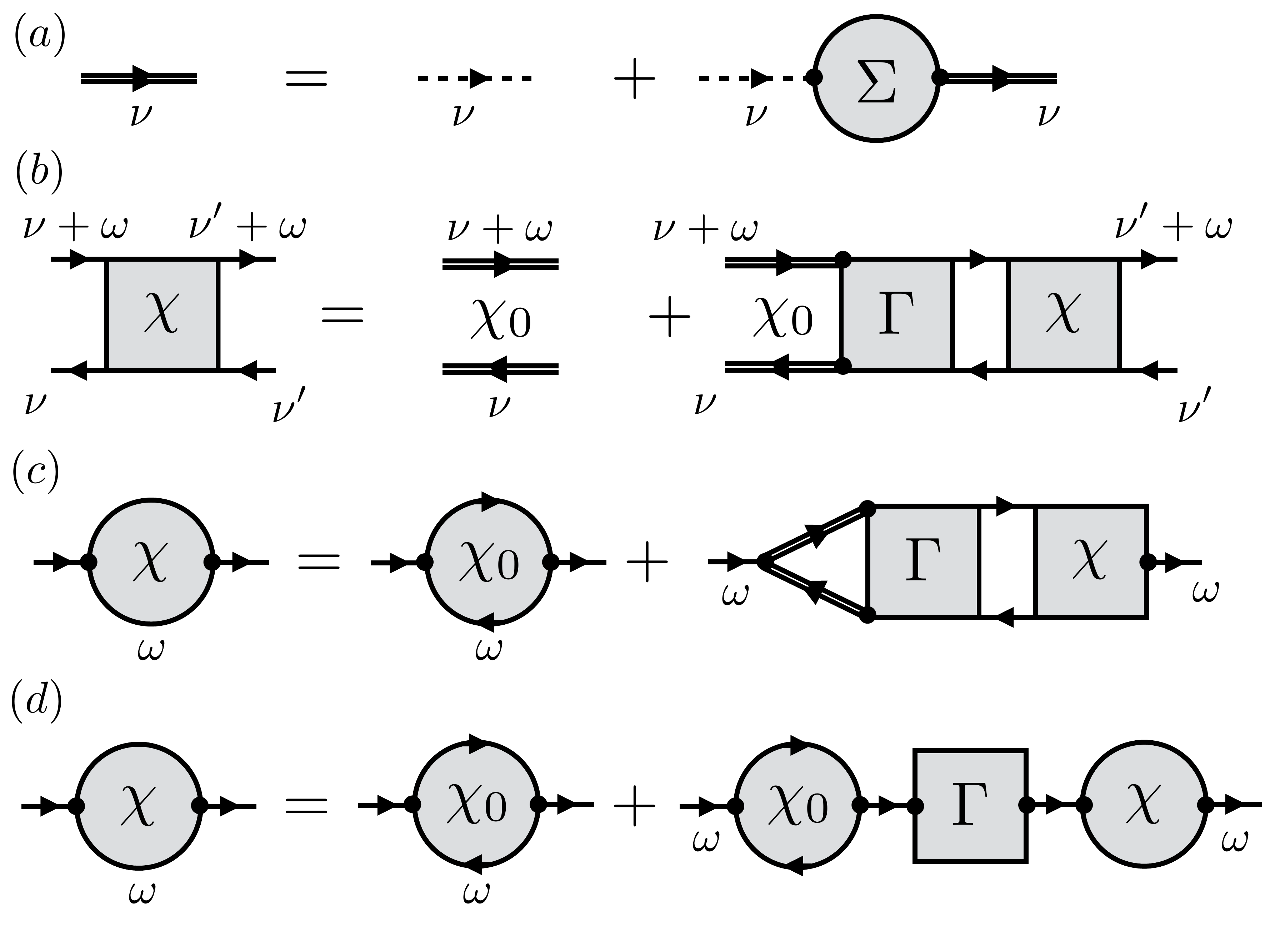}}
\caption{Diagrams used in cluster perturbation theory: (a) the Dyson equation, (b) the Bethe-Salpeter equation for the generalized four-point susceptibility, (c) the Bethe-Salpeter equation for the two-point susceptibility and (d) the analogue of the Dyson equation for the two-point susceptibility. The four-point vertex, $\Gamma(i\nu,i\nu',i\omega)$, is approximated by the two-point vertex, $\Gamma(i\omega)$.  Dashed lines represent the non-interacting Green's function $G_0$, while double lines represent the interacting Green's function $G$. }
\label{fig_Diagrams}
\end{figure}

CPT couples clusters perturbatively in the inter-cluster hopping to form the full lattice. Another way of deriving the CPT equations for the full lattice Green's function is through the Dyson equation:
$$\bm{G}(\tilde{\bf{k}},i\nu) = \bm{G}_0(\tilde{\bf{k}},i\nu) + \bm{G}_0(\tilde{\bf{k}},i\nu)\bm{\Sigma}(\tilde{\bf{k}},i\nu)\bm{G}(\tilde{\bf{k}},i\nu) \,. $$
\noindent
Here $\bm{G}$ is the connected Green's function, $\bm{G}_0$ is the non-interacting Green's function, and $\bm{\Sigma}$ is the sum of all one-particle-irreducible diagrams. All quantities are matrices in the cluster sites.  Fig.~\ref{fig_Diagrams}a depicts the Dyson equation diagrammatically for later comparison. A similar relation holds for the corresponding quantities on a cluster:
$$\bm{G}_{\text{c}}(i\nu) = \bm{G}_{0,\text{c}}(i\nu) + \bm{G}_{0,\text{c}}(i\nu)\bm{\Sigma}_{\text{c}}(i\nu)\bm{G}_{\text{c}}(i\nu). $$
These relations allow for an approximate solution for the CPT Green's function.

To derive the CPT approximation to the Green's function we assume:
$$ \bm{\Sigma}(\tilde{\bf{k}},i\nu) \approx \bm{\Sigma}_\text{c}(i\nu).$$
This is the central approximation to one-particle CPT that allows us to 
solve for the lattice connected Green's Function, yielding:
\begin{align}
 \bm{G}_{\text{CPT}}^{-1}(\tilde{{\bf{k}}},i\nu) =  \bm{G}_{0}^{-1}(\tilde{{\bf{k}}},i\nu)  +\bm{G}_\text{c}^{-1}(i\nu)
 -\bm{G}_{0,\text{c}}^{-1}(i\nu).
\label{eq_G_CPT}
\end{align}
This CPT approximation to the Green's function can be computed numerically in a straightforward fashion because the mixed representation allows us to insert an approximation for the self-energy derived from numerically solving the small sized cluster.  Appendices \ref{app_twopoint} and \ref{app_CPT} detail how Eq.~\ref{eq_G_CPT} is computed numerically with exact diagonalization.

We can use $G_{\text{CPT}}$ to compute one-particle correlation functions in momentum space to bridge small cluster results (with a momentum space mesh) and large systems (with a continuous momentum space). To express the Green's function in momentum space we take a Fourier transform and ignore the off-diagonal cluster momenta to restore translational invariance:
\begin{align}
G({\bf{k}},i\nu) \approx \frac{1}{V_d}\sum_{a,b}e^{-i{\bf{k}}\cdot({{\bf r}_a}-{{\bf r}_b})}[G_
\text{CPT}({\bf{k}},i\nu)]_{ab}, 
\label{eq_G_CPT_q}
\end{align}
where $V_d = L^d$ is the volume of the $d$-dimensional cluster and we have set $\tilde{\bm{k}} = \bm{k}$ since $\bm{K}$ is periodic.
It is important to note that this approximation for  $ G({\bf{k}},i\nu)$ is equivalent to exact results at $U=0$ and $t=0$. For $U=0$ there is no self-energy and we recover the non-interacting Green's function. In the atomic limit, $t=0$, we retrieve the Green's function for just a single site, where  $G_{0,\text{CPT}} = G_{0,\text{c}}$, so then $G = G_\text{c}$. Additionally, in the limit $L\rightarrow\infty$, CPT is equivalent to the full lattice. CPT systematically approaches the thermodynamic limit with increasing cluster size $V_d$. Using this structure we can construct an analogous procedure for two-particle correlation functions.

\section{Two-Particle Cluster Perturbation Theory}
\label{sec_twopart}

\begin{figure}
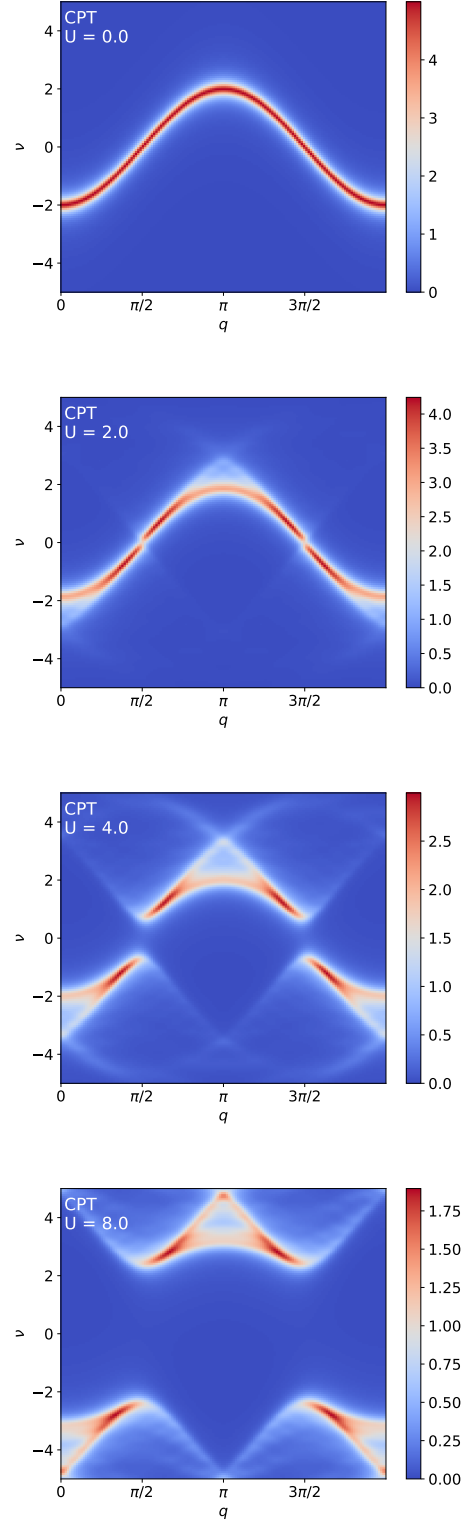

	\centering
	\subfloat{\includegraphics[scale=.4]{{{Figures/L16Figs/SpectralFunction/AL16U0.0P144eps0.2}}}}\\
	\subfloat{\includegraphics[scale=.4]{{{Figures/L16Figs/SpectralFunction/AL16U2.0P144eps0.2}}}}\\
	\subfloat{\includegraphics[scale=.4]{{{Figures/L16Figs/SpectralFunction/AL16U4.0P144eps0.2}}}}\\
	\subfloat{\includegraphics[scale=.4]{{{Figures/L16Figs/SpectralFunction/AL16U8.0P144eps0.2}}}}\\
	\caption{CPT approximation (Eq.~\ref{eq_G_CPT_q}) to the Green's Function, $-\operatorname{Im}G(k,\nu+i\eta)$, for the one-dimensional Hubbard model at half filling plotted as a function of wavevector $k$. We choose a cluster size $L = 16$, small broadening parameter $\eta = 0.2$, and plot $N_p = 144$ points for $k$ and $\nu$. From top to bottom we have Hubbard interaction strengths: $U=0,2,4,$ and $8$.
	 }
	\label{fig_CPTG}
\end{figure}

\begin{figure*}
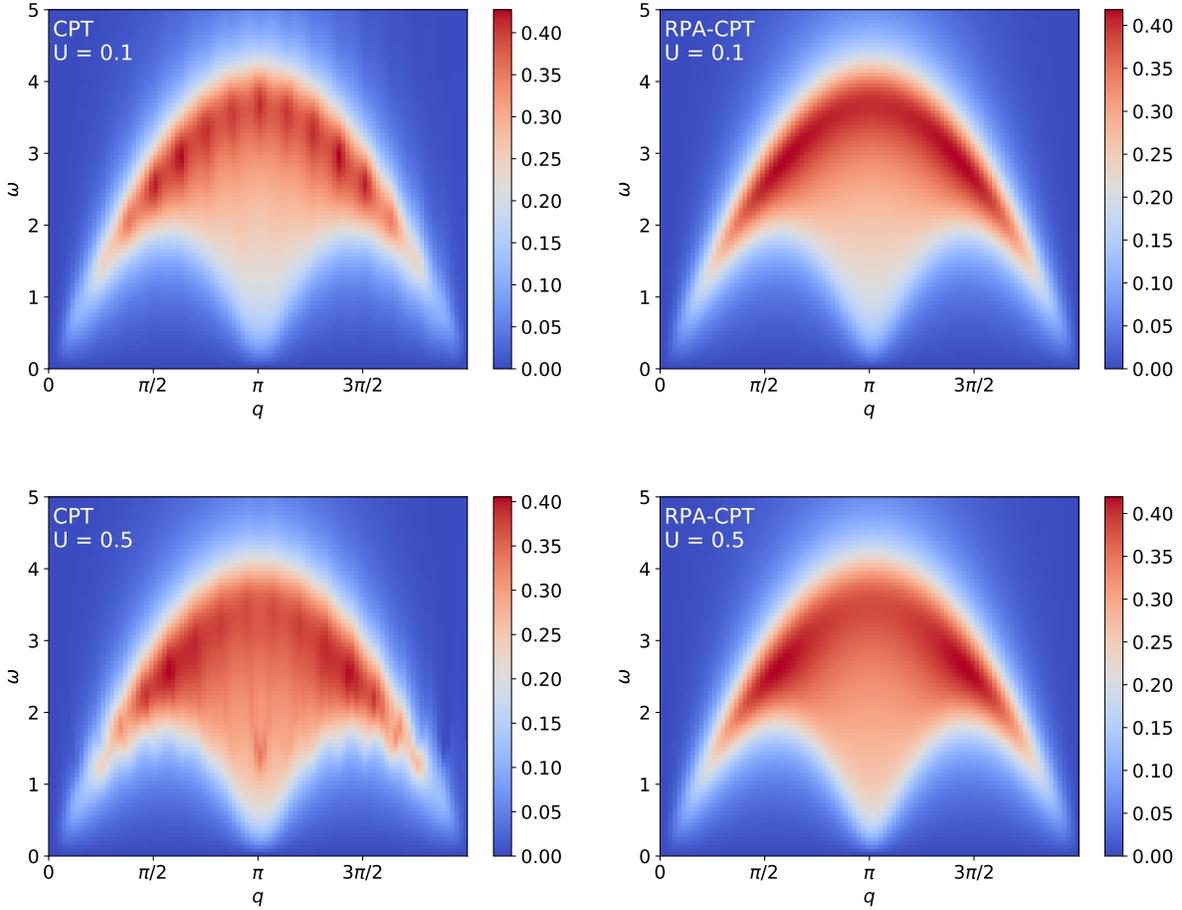

	\centering
	\subfloat{\includegraphics[scale=.5]{{{Figures/L16Figs/CPT/imchiGL16U0.1P96eps0.5}}}}
	\subfloat{\includegraphics[scale=.5]{{{Figures/L16Figs/RPA/imchiRPAL16U0.1P96eps0.5}}}}\\
	\subfloat{\includegraphics[scale=.5]{{{Figures/L16Figs/CPT/imchiGL16U0.5P96eps0.5}}}}
	\subfloat{\includegraphics[scale=.5]{{{Figures/L16Figs/RPA/imchiRPAL16U0.5P96eps0.5}}}}
	\caption{ Left: CPT approximation (Eq.~\ref{eq_chiq}) to the spin susceptibility, $\operatorname{Im}\chi(q,\omega+i\eta)$ for the one-dimensional Hubbard model at half filling plotted as a function of wavevector. We choose $L=16, \eta = 0.5$, and $N_p=96$. Right: The same but for the RPA-CPT approximation, Eq.~\ref{eq_cpt_rpa}. The interaction is chosen to be weak, $U=0.1$ and $U=0.5$, (top and bottom respectively) so that the RPA-CPT is accurate. The oscillating lines are a numerical artifact discussed in Sec.~\ref{app_numerical}
	 }
	\label{fig_lowU}
\end{figure*}

\begin{figure*}
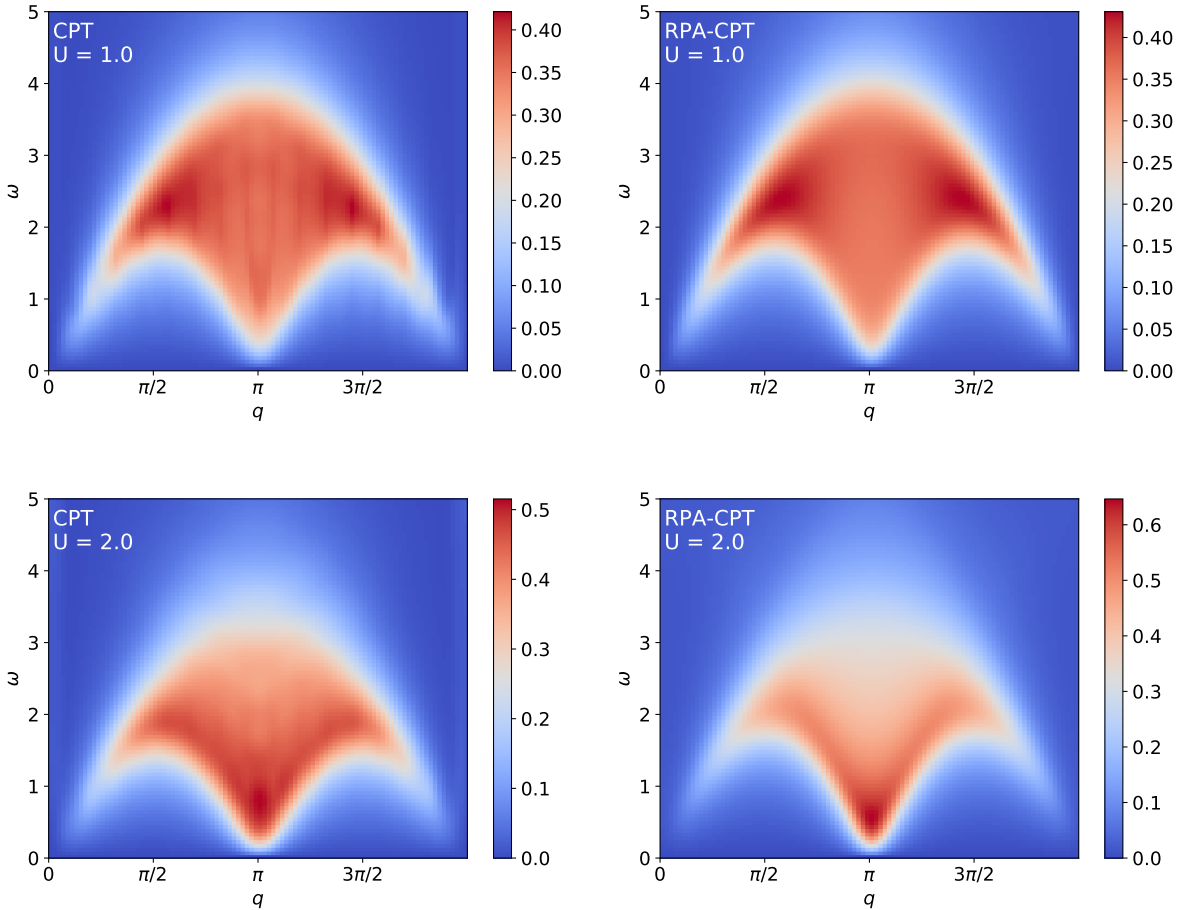

	\centering
	\subfloat{\includegraphics[scale=.5]{{{Figures/L16Figs/CPT/imchiGL16U1.0P96eps0.5}}}}
	\subfloat{\includegraphics[scale=.5]{{{Figures/L16Figs/RPA/imchiRPAL16U1.0P96eps0.5}}}}\\
	\subfloat{\includegraphics[scale=.5]{{{Figures/L16Figs/CPT/imchiGL16U2.0P96eps0.5}}}}
	\subfloat{\includegraphics[scale=.5]{{{Figures/L16Figs/RPA/imchiRPAL16U2.0P96eps0.5}}}}\\
  \caption{ Same as Fig.~\ref{fig_lowU} but for an intermediate interaction strength, $U=1$ (top) and $U=2$ (bottom). The CPT (left) and RPA-CPT (right) agree well up to $U=2$ where deviations between the methods start to appear. 
	 }
	\label{fig_intermediateU}
\end{figure*}

\begin{figure*}
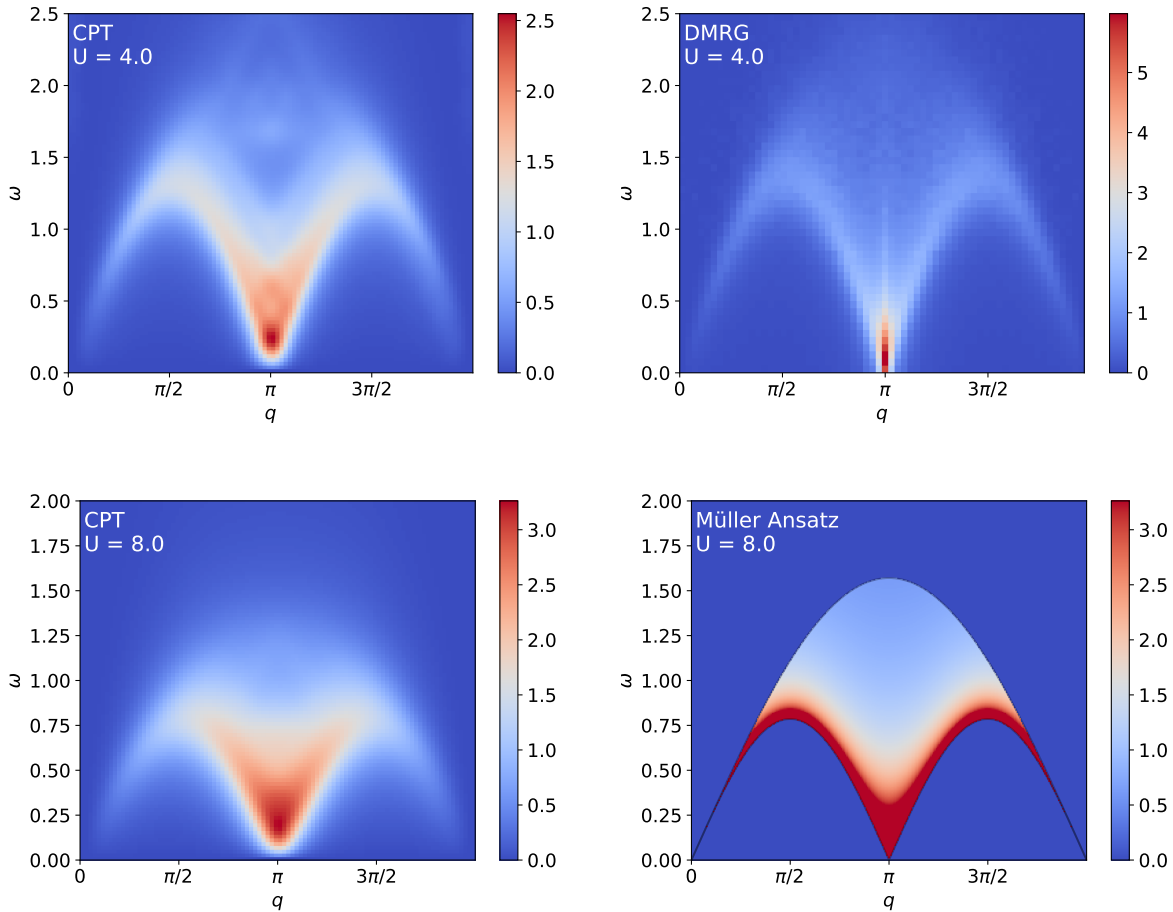

\centering
	\subfloat{\includegraphics[scale=.5]{{{Figures/L16Figs/CPT/imchiGL16U4.0P96eps0.2}}}}
  \subfloat{\includegraphics[scale=.5]{{{Figures/L16Figs/imchiDMRG}}}}\\ 
	\subfloat{\includegraphics[scale=.5]{{{Figures/L16Figs/CPT/imchiGL16U8.0P96eps0.2}}}}
	\subfloat{\includegraphics[scale=.5]{{{Figures/imchiMAL16U8.0P288eps0.2}}}}
  \caption{Left: The same as Fig.~\ref{fig_lowU} but for larger interaction strengths, $U=4$ and $U=8$, and with $\eta = 0.2$. Right (top): DMRG result for a 64 site chain and $U=4$.  Here we see that the CPT method broadens the peak at $q=\pi$ in comparison to DMRG.   Right (bottom): $\text{M{\"u}ller}$ estimate, Eq.~\ref{eq_muller}, for $U=8$ with a cutoff intensity such that $\text{max}[\chi_{\text{M{\"u}ller}}(q,\omega)] = \text{max}[\operatorname{Im}{\chi}_\text{CPT}(q,\omega+i\eta)]$. }
\label{fig_U4}
\end{figure*}

We now apply the logic of the one-particle CPT formalism reviewed in Sec.~\ref{sec_onepart} to two-particle correlation functions.  A general two-particle correlation function in the particle-hole channel is given by: 
\begin{align}
\nonumber
\chi_{\alpha\alpha'\beta\beta'}(i\omega,i\nu,i\nu') &=\int_0^\beta \!\!d\bm{\tau} e^{-i\omega\tau_1}e^{(i\nu+i\omega)\tau_2}e^{-(i\nu'+i\omega)\tau_3}\\
&\times \langle T_{\bm{\tau}}c_{\alpha}^\dagger(\tau_1)c_{\alpha'}(\tau_2)c_{\beta}^\dagger(\tau_3)c_{\beta'}(0) \rangle,
\nonumber
\end{align}

\noindent
where $\bm{\tau} = (\tau_1,\tau_2,\tau_3)$. In general there are four independent frequencies, but due to energy conservation, they can be reduced to only three: $i\omega, i\nu$, and $i\nu'$. Where $i\omega$ is the bosonic transfer Matsubara frequency and $i\nu$ and $i\nu'$ are fermionic Matsubara frequencies. To approximate a two-particle correlation function using CPT we start with the Bethe-Salpeter equation (represented diagramatically in Fig.~\ref{fig_Diagrams}b):

\begin{align}
&\bm{\chi}(i\omega,i\nu,i\nu') = \bm{\chi}_0(i\omega,i\nu,i\nu')\delta_{i\nu,i\nu'}\\
&+\sum_{i\nu'',i\nu'''}\bm{\chi}_0(i\omega,i\nu,i\nu'')\bm{\Gamma}(i\omega,i\nu'',i\nu''')\bm{\chi}(i\omega,i\nu''',i\nu') \nonumber
\end{align} 
where,
$$\bm{\chi}_0(i\omega,i\nu,i\nu') = \bm{G}(i\nu+i\omega)\bm{G}(i\nu), $$

\noindent
and all quantities are matrices in the site indices. The four-point vertex $\bm{\Gamma}$ is the sum of all scattering diagrams that are irreducible in the particle-hole channel.  The full Bethe-Salpeter equation is expensive to evaluate for any arbitrary two-particle correlation function because it is a rank-4 tensor and has 3 independent frequencies. To connect with inelastic neutron scattering experiments and keep the discussion tangible, we focus on the spin susceptibility defined as:

\begin{align}
\chi_{ij}(i\omega) &= \int_0^{1/k_BT} d\tau e^{i\omega\tau}\langle T_{\tau}S_i^+(\tau)S_j^-(0)\rangle \label{eq_chi_ij}\\
&= \int_0^{1/k_BT} d\tau e^{i\omega\tau} \langle T_{\tau}c_{i,\uparrow}^\dagger(\tau)c_{i,\downarrow}(\tau)c_{j,\downarrow}^\dagger(0)c_{j,\uparrow}(0)\rangle \nonumber
\end{align}

\noindent
where the spin raising and lowering operators are given by $S_i^+(\tau)=c_{i,\uparrow}^\dagger(\tau)c_{i,\downarrow}(\tau)$ and $S_{j}^-(\tau)=c_{j,\downarrow}^\dagger(\tau)c_{j,\uparrow}(\tau)$, respectively. Note that here $\tau_1 = \tau_2 = \tau$ and $\tau_3 = 0$. In other words, we join the outer legs in Fig.~\ref{fig_Diagrams}b making $i\omega$ the external frequency and $i\nu$ and $i\nu'$ internal frequencies to produce Fig.~\ref{fig_Diagrams}c.

As stated previously, evaluating the full two-particle correlation function, and hence the full 4-point vertex $\bm{\Gamma}$, is expensive. To reduce the computational cost, we neglect the dependence of $\bm \Gamma$ on the internal frequencies (and momentum), i.e., 
\begin{align}\label{eq:GammaApprox}
\bm{\Gamma}(i\omega,i\nu,i\nu') \approx \bm{\Gamma}(i\omega)\,.
\end{align}
This simplification was argued in Ref.~\onlinecite{Brehm2010} to provide a good approximation for the magnetic particle-hole channel, for which the dominant correlations have very little internal frequency (and momentum) dependence. This approximation then gives a Dyson-like two-leg Bethe-Salpeter equation depicted diagramatically in Fig.~\ref{fig_Diagrams}d and given by:
\begin{subequations}
\begin{align}
\bm{\chi}(\tilde{\bf{q}},i\omega) &= \bm{\chi}_0(\tilde{\bf{q}},i\omega) + \bm{\chi}_0(\tilde{\bf{q}},i\omega)\bm{\Gamma}(\tilde{\bf{q}},i\omega)\bm{\chi}(\tilde{\bf{q}},i\omega),
\label{eq_twoptBSE}
\\
\bm{\chi}_{\text{c}}(i\omega) &= \bm{\chi}_{0,\text{c}}(i\omega) + \bm{\chi}_{0,\text{c}}(i\omega)\bm{\Gamma}_{\text{c}}(i\omega)\bm{\chi}_{\text{c}}(i\omega)
\label{eq_twoptBSEc}
\end{align}
\end{subequations}
for the lattice and cluster respectively. 

We can now apply the CPT procedure outlined in Sec.~\ref{sec_onepart} to the spin susceptibility. Using the above two equations for the lattice and cluster spin susceptibilities along with the central assumption of our work:
\begin{align}
 \bm{\Gamma}(i\nu,i\nu',i\omega) \approx \bm{\Gamma}(i\omega) \approx \bm{\Gamma}_\text{c}(i\omega), 
 \label{eq_gamma_app}
\end{align}
\noindent
we can derive a CPT approximate spin susceptibility for the lattice:
\begin{align}
 \bm{\chi}_{\text{CPT}}^{-1}(\tilde{{\bf{q}}},i\omega) = \bm{\chi}_{0,\text{CPT}}^{-1}(\tilde{{\bf{q}}},i\omega) + \bm{\chi}_\text{c}^{-1}(i\omega)
 -\bm{\chi}_{0,\text{c}}^{-1}(i\omega),
\label{eq_chiqCPT}
\end{align} 
\noindent
where $\bm{\chi}_{0,\text{CPT}}$ is the dressed bubble diagram using the CPT Green's function, Eq.~\ref{eq_G_CPT}. The appendix ~\ref{app_bubble} details how to compute the bubble diagram. As a final step we make contact with the momentum space spin susceptibility in the full lattice:
\begin{align}
\chi({\bf{q}},i\omega) \approx   \frac{1}{V_d}\sum_{a,b}e^{-i{\bf{q}}\cdot({\bf r}_a-{\bf r_b})}[\chi_
\text{CPT}({\bf{q}},i\omega)]_{ab}. 
\label{eq_chiq}
\end{align}
This approximation for $\chi$ can now be compared with experimental data continuous in $q$.  In principle this procedure can be used similarly for other two-particle correlation functions. 
However, depending on the channel and the structure of the dominant correlations in that channel, one may have to dress the irreducible four-point vertex with appropriate form-factors in Eq.~\ref{eq:GammaApprox} that describe the structure of the leading correlations in that channel. This generalization will be discussed in future work. 

The above scheme leads to an approximation for the spin susceptibility which is exact in certain limits.  As with CPT for the single-particle Green's function, we retrieve the exact results in the two limits for the Hubbard model, $U=0$ and $t=0$. For $U=0$, there is no self-energy, the irreducible vertex is zero, and we recover the non-interacting $\chi$.  In the $t=0$ limit we are once again restricted to the single site and $\chi_{0,\text{CPT}} = \chi_{0,\text{c}}$ giving $\chi = \chi_\text{cluster}$.  Additionally, the CPT approximation for the spin susceptibility is systematically improved with increasing cluster size $L$, approaching the exact result as $L\rightarrow\infty$. The applicability of the method away from these points depends on the model.  In the next section we benchmark the approximation on a model where essentially exact results are accessible.

\section{Application to the One-Dimensional Hubbard Model at Half-filling}
\label{sec_1dhubbard}

We now test the accuracy and viability of the CPT approximation to the spin susceptibility on a specific model.  The method applies generally to $H$ in any dimension but here we focus on a model where exact results are known. We consider the single-band $d=1$ Hubbard model:
$$ H_{\text{1D}} = -t\sum_{\langle  i,j \rangle, \sigma\in \uparrow,\downarrow} c_{ i,\sigma}^\dagger c_{ j,\sigma} + U\sum_{i} n_{i\uparrow}n_{ i\downarrow} - \mu\sum_{ i,\sigma} n_{ i\sigma}. $$
The 1D Hubbard model has been studied extensively \cite{1385i}.  Exact results are known for single-particle correlation functions in the ground state.  The only known exact results for the spin susceptibility involve scaling near poles or extreme limits (weak and strong interaction).  DMRG offers an accurate method for comparison of the spectral function at intermediate interaction strengths \cite{Nocera2016,Nocera2016a,Nocera2018}.   In this section we focus on half filling (density $n=1$, where $n\equiv\sum_{i,\sigma}\langle n_{i,\sigma}\rangle$) obtained by setting $\mu=U/2$ (Appendix~\ref{app_chemical_potential} discusses how to estimate $\mu$ away from half filling).  We also work at zero temperature, and use units with $t=\hbar=1$. We also set the lattice spacing constant to unity, $a_0 \equiv 1$.

First, we test our numerical implementation for the single-particle CPT.  By computing the spectral function we reproduce work in Ref.~\onlinecite{Senechal2000}, which has been benchmarked against exact results.  We find that the CPT results for the Green's function in Fig.~\ref{fig_CPTG} also agree with the time-dependent density matrix renormalization group results for the Green's function in Ref.~\onlinecite{Nocera2018}.

\begin{figure*}[t]
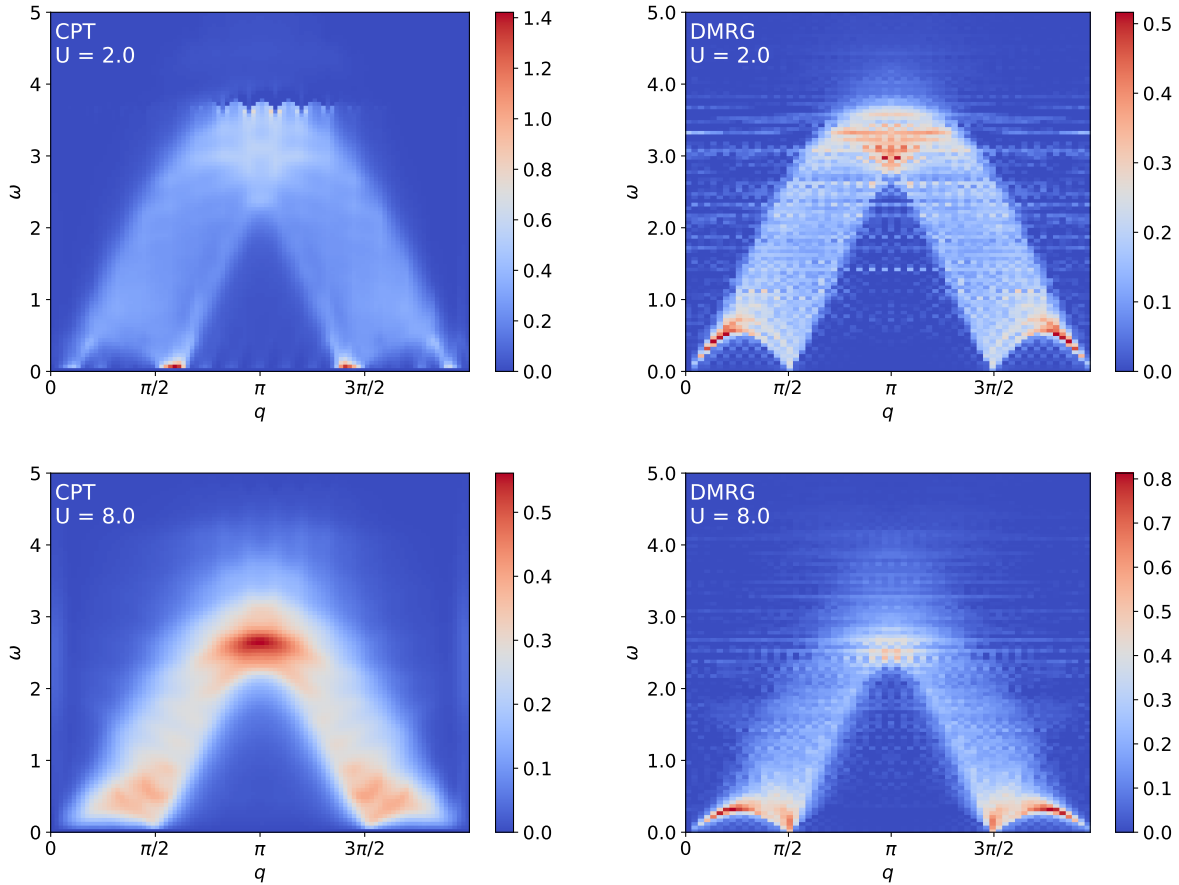

  \centering
  \includegraphics[scale=.5]{{{Figures/Off-Half/CPT/imchiGL16U2.0P96eps0.2N4}}}
  \includegraphics[scale=.5]{{{Figures/Off-Half/DMRG/imchiDMRGN4U2}}}
  \includegraphics[scale=.5]{{{Figures/Off-Half/CPT/imchiGL16U8.0P96eps0.2N4}}}
  \includegraphics[scale=.5]{{{Figures/Off-Half/DMRG/imchiDMRGN4U8}}}
  \caption{ Left: CPT approximation (Eq.~\ref{eq_chiq}) to the spin susceptibility, $\operatorname{Im}\chi(q,\omega+i\eta)$ for the one-dimensional Hubbard model at one-quarter filling (density $n=1/2$) plotted as a function of wavevector. We choose $L=16, \eta = 0.2$, and $N_p=96$.  The chemical potentials used are listed in Sec.~\ref{app_chemical_potential}. Numerical artifacts in CPT are due to pole mismatches discussed in  Secs.~\ref{app_chi} and \ref{app_numerical}.  Right: DMRG result for a 64 site chain at the same filling with $\eta = .05$.  }
  \label{fig_chi_comp_quarter}
\end{figure*}

We now move beyond single particle correlation functions to compare CPT results for the spin susceptibility $\chi$ with known results.  In the low U limit, we can accurately use perturbation theory on $ H_{\text{1D}}$.  Here the RPA should be reasonably accurate, and leads to:
\begin{align}
 \chi_{\text{RPA-CPT}}(q,i\omega) = \frac{\chi_{0,\text{CPT}}(q,i\omega)}{1 - U\chi_{0,\text{CPT}}(q,i\omega)}; U\ll t.
 \label{eq_cpt_rpa}
 \end{align}
 Here, $\chi_{0,\text{CPT}}$ is the bare susceptibility (bubble diagram) computed with the dressed CPT Green's function $G_{\rm CPT}$ in Eq.~\ref{eq_G_CPT}. We call this approximation RPA-CPT because it amounts to replacing the full vertex $\bm\Gamma$ in Eq.~\ref{eq_twoptBSE} by its leading order (RPA) approximation $U$.
 
Figures~\ref{fig_lowU} and \ref{fig_intermediateU} compare the CPT to the RPA approximation.   Here we see that both methods agree at low $U$.  But the CPT results also show oscillations which are numerical artifacts. The 1D Hubbard model is SU(2)-symmetric; therefore, the cluster spin susceptibility, $\bm{\chi}_c(\omega +i\eta)$  is non-invertible for all $\omega$'s\cite{Filor2014a}. The appendices \ref{app_chi} and \ref{app_numerical} explicitly detail how we evaluate Eq.~\ref{eq_chiqCPT}, and how we alleviate finite-size numerical issues.  These effects arise for a small broadening parameter $\eta$.  For comparison with experiments, the small $\eta$ limit may not be needed due to intrinsic experimental broadening in measures of the spin susceptibility.  The appendix (Sec.~\ref{app_vertex}) shows the frequency dependence of the CPT vertex function ${\bm \Gamma}(\omega)$. We find that for small $U$ and for $\omega$ below $\approx4$,  ${\bm \Gamma}(\omega) \approx U$, i.e. the RPA approximation in Eq.~\ref{eq_cpt_rpa} and CPT agree reasonably well.

We now turn to comparisons at larger $U$. The top panels in Fig.~\ref{fig_U4} compare our CPT results for $\chi$, Eq.~\ref{eq_chiq}, with DMRG \cite{Nocera2016,Nocera2016a,Nocera2018}.  Here the DMRG system size, 64 sites, is converged enough to approximate the thermodynamic limit.  We see that DMRG produces a narrower/taller peak for $\omega<0.05$ and $q=\pi$, otherwise all of the other qualitative features appear to be the same.  The limited CPT cluster size (16 sites chosen here), broadens the peak near $q=\pi$ for the left plot because there are fewer $q$ points to sample. 

In the very large $U$ limit the 1D Hubbard model maps to the isotropic Heisenberg spin chain with spin-spin interaction $ J\simeq 4t^2/U$.  For $U=4$ and $U=8$, DMRG calculations \cite{Nocera2016a} of $\chi(q,\omega)$ on the Hubbard model show close agreement with the Heisenberg model. In the Heisenberg limit, a well known phenomenological estimate for the spin susceptibility can be used for comparison to CPT \cite{Muller1981}:
\begin{align}
\chi_{\text{M{\"u}ller}}(q,\omega) \approx \frac{\Theta[\omega-\omega_L(q)]\Theta[\omega_U(q)-\omega]}{\sqrt{\omega^2-\omega_L^2(q)}}; U\gg t,
\label{eq_muller}
\end{align}
where $\Theta$ is the step function.  The lower energy branch is given by $\omega_L(q) = (\pi/2)J|\sin(qa_0)|$ and the upper energy branch is given by $\omega_U(q) = \pi J|\sin(qa_0/2)| $, thus reproducing the Cloiseaux-Pearson relations.  Eq.~\ref{eq_muller} yields accurate results in comparison to the exact results for the Heisenberg model \cite{Bougourzi1996} for the large interaction limit because it was chosen to match small system size Bethe-Ansatz results while respecting sum rules and the Cloiseaux-Pearson relations.  Eq.~\ref{eq_muller} deviates from exact results on the upper boundary of the Cloiseaux-Pearson relation because it incorrectly predicts a step in the upper boundary.

\begin{figure*}[t]
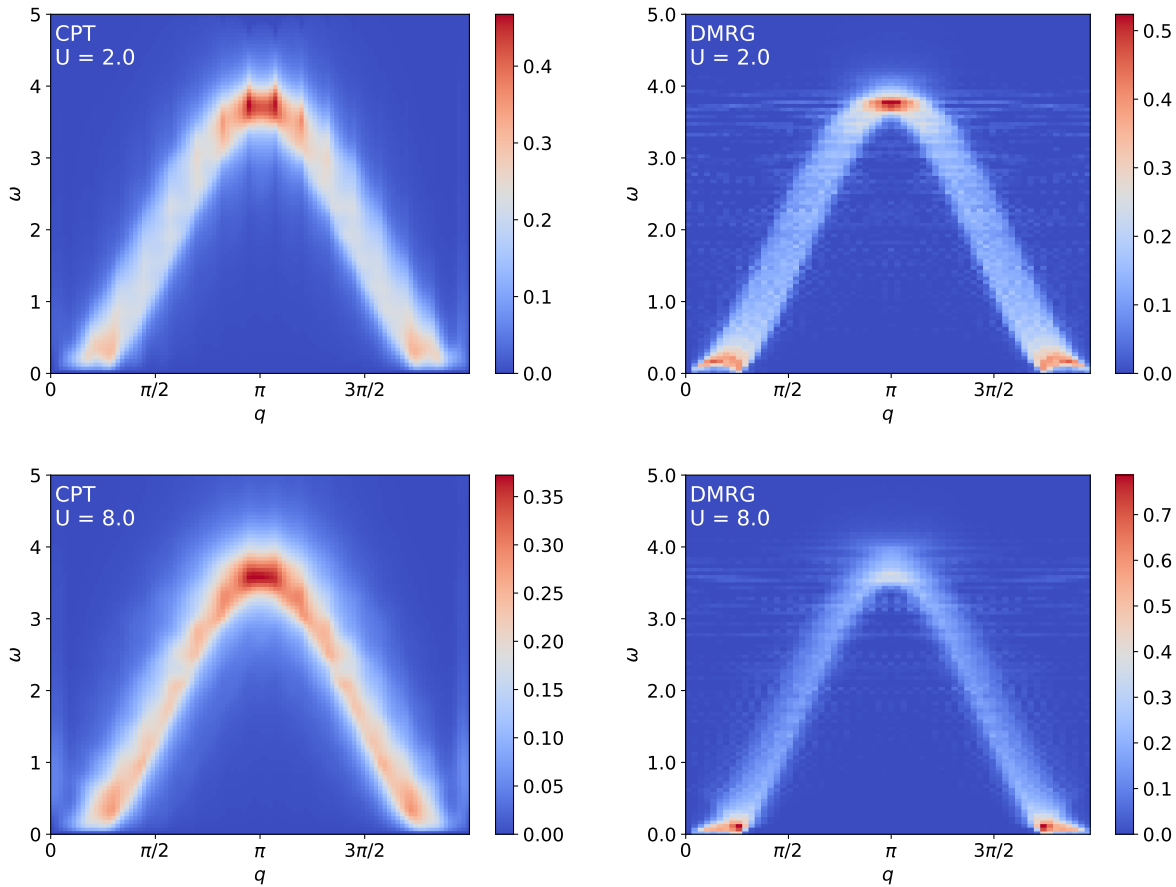

  \centering
  \includegraphics[scale=.5]{{{Figures/Off-Half/CPT/imchiGL16U2.0P96eps0.2N8}}}
  \includegraphics[scale=.5]{{{Figures/Off-Half/DMRG/imchiDMRGN8U2}}}
  \includegraphics[scale=.5]{{{Figures/Off-Half/CPT/imchiGL16U8.0P96eps0.2N8}}}
  \includegraphics[scale=.5]{{{Figures/Off-Half/DMRG/imchiDMRGN8U8}}}
  \caption{Same as Fig. \ref{fig_chi_comp_quarter} except for one-eighth filling (density $n = 1/4$).}
  \label{fig_chi_comp_oct}
\end{figure*}

Figure~\ref{fig_U4} compares our CPT results for $\chi$ with the $\text{M{\"u}ller}$ estimate.  We find that CPT better approximates the $\text{M{\"u}ller}$ estimate at larger $U$.  This is reasonable because the $\text{M{\"u}ller}$ estimate was derived in the Heisenberg limit.  But we do see notable differences even at $U=8t$.  For example, there is less weighting in $\chi$ at large $\omega$, near $\omega\sim 1-1.5t$.  The shifting of spectral weight from the upper spinon branch to the lower spinon branch is as expected\cite{Bhaseen2005}.  The larger $U$ enhances the peak at $q=\pi$ and is evidence for strong antiferromagnetic correlations.

The CPT graphs in Fig.~\ref{fig_U4} shows a small gap at $q=\pi$.  This is due to the finite size of the cluster.  The gap here should vanish in the thermodynamic limit \cite{Karbach1997}.  The Appendix (Sec.~\ref{app_scaling_gap}) discusses extrapolations of the gap as a function of cluster size.  We find that the gap tends to diminish as $L\rightarrow\infty$.  But we find that finite size effects do not allow a precise extrapolation to zero.

So far we have found that CPT for two-particle correlation functions offers a good approximation for the 1D Hubbard model at half filling.  CPT reduces to exact results for the spin susceptibility in weak and strong interaction limits.  From Figs.~\ref{fig_lowU}-~\ref{fig_U4}, we see that for a wide range of interaction strengths CPT for the spin-susceptibility agrees reasonably well with benchmark methods.  These are nontrivial tests because quantum fluctuations in the spin sector are strong at half filling in 1D.

\section{Application to the One-Dimensional Hubbard Model away from  half-filling}
\label{sec_awayhalf}

The results in the previous sections show that, at half filling, CPT estimates of the spin susceptibility compare reasonably well against benchmarks even with strong quantum fluctuations in the spin sector. We now turn our attention to results away from half-filling where we also expect strong charge fluctuations.  At fillings below one-half we have a gapless state where both charge and spin fluctuations are strong.  We compare the CPT results to the DMRG results for two different fillings ($1/4$ and $1/8$) and two different interaction strengths ($U = 2$ and $U = 8$).  Since CPT is an approximation we expect to find disagreement at intermediate $U$, and indeed, we do find quantitative disagreement with DMRG at these lower fillings, as expected. But we also find regimes where CPT gives a good qualitative comparison with DMRG for the spin susceptibility.  

To use CPT away from half filling we must estimate the chemical potential.  Sec.~\ref{app_chemical_potential} describes the procedure by which we  determine $\mu$ at fixed density.  The results presented here use the chemical potentials computed in Sec.~\ref{app_chemical_potential}. 

Figures~\ref{fig_chi_comp_quarter} and \ref{fig_chi_comp_oct} compare the spin susceptibilities for one-quarter and one-eigth filling, respectively.  Results are presented for both CPT and DMRG.  As in previous sections, we choose a relatively small cluster size for CPT ($L=16$) and 64 sites for DMRG.  For comparison purposes we assume that the DMRG results are a good approximation to the exact results in the thermodynamic limit. 

Figures~\ref{fig_chi_comp_quarter} and \ref{fig_chi_comp_oct} show significant differences in the spectral weight when comparing CPT and DMRG. For $U=2$ (top row, Fig.~\ref{fig_chi_comp_quarter}), we see at low $\omega$ that CPT correctly predicts low energy excitations near $q=0$ and $\pi/2$.  However, the CPT results show stronger weight near $q=\pi/2$, whereas DMRG indicates enhanced weight near $q=0$.  The $U=8$ results (bottom rows of Fig.~\ref{fig_chi_comp_quarter} and \ref{fig_chi_comp_oct}) also show discrepancy in the spectral weight.  The CPT results show strong weight near $\omega\sim3$.  DMRG, by contrast, shows the largest weight near $\omega\sim0$.

\section{Summary}
\label{sec_summary}

CPT is a simple and economic method to compute the momentum-resolved Green's function for Hubbard models with local interactions. CPT with exact diagonalization is useful for exploring parameter space because it is less complex than competing methods such as quantum Monte Carlo, density matrix renormalization group, or dynamical mean field theory. We find that the CPT-extension to a higher order correlation function, the transverse spin-susceptibility, allows a relatively economic and accurate implementation in the 1D Hubbard model.  Our results also suggest that the method can be applied to other more sophisticated Hubbard models where clusters offer reasonable approximations to bulk physics.  We note, however, that CPT itself is limited to regimes with low entanglement. CPT estimates of one and two-particle correlation functions will probably offer rather poor approximations in models with higher degrees of entanglement, e.g., the 2D Hubbard model at intermediate interaction strengths.  

In our benchmarking against exact results for the 1D Hubbard model we have found that the CPT method shows deviations from expected behavior due to two primary issues.  The first issue is the finite size of the cluster, which leads to a finite gap at $q=\pi$. Finite-size scaling is needed to extrapolate to a zero gap and make more accurate predictions for the thermodynamic limit.  The second issue arose from pole mismatches in approximations to $\chi$ that led to numerical artifacts (Secs.~\ref{app_chi} and ~\ref{app_numerical}) particularly when our broadening parameter was small.  A general solution could be to construct a Lehman representation of $\chi_{\text{CPT}}$.  This would be useful for a more numerically controlled CPT-based method when small broadening parameters are needed. 

\begin{acknowledgments}
V.W.S. and P.T.R. acknowledge support from AFOSR (FA9550-18-1-0505) and ARO (W911NF-16-1-0182). The work of T.A.M and G.A. was supported by the Scientific Discovery through Advanced Computing (SciDAC) program funded by U.S. Department of Energy, Office of Science, Advanced Scientific Computing Research and Basic Energy Sciences, Division of Materials Sciences and Engineering.
\end{acknowledgments}

\appendix
\renewcommand\thefigure{\thesection\arabic{figure}}

\section{$Q$-Matrix representation for two-point correlation functions}
\label{app_twopoint}

Here we discuss approximation of  two-point dynamical imaginary time correlation functions 
$\braket{T_\tau \mathcal{O}_{\alpha}(\tau)\mathcal{O}_{\beta}^\dagger(0)}$ 
on a single cluster.  Ordinarily the use of the Lanczos method leads to a large number of poles. In this section we show how we reduce calculation of such correlators for a single cluster to a representation in terms of ``$Q$-Matrices".  The new representation allows use of band Lanczos\cite{Meyer1989} where only a single set of poles arises for each element. 

To be specific we focus on  the Green's function $G = -\braket{T_\tau c_{a}(\tau)c_{b}^\dagger(0)}$ although the same procedure works for $\chi = -\braket{T_\tau S_{a}^+(\tau)S_{b}^-(0)}$. We first pass to the Lehman representation.  After constructing the thermal average of the Green's function, we can insert an eigenstate basis and rewrite the imaginary time operators only considering $\tau>0$.  After a Fourier transform into frequency space, using the (anti)periodicity $e^{i\nu k_BT} = \mp1$ for fermions and bosons respectively, and simplifying terms, we have:
\begin{align}
G_{\alpha\beta}(i\nu) = \frac{1}{Z}\sum_{n,m}\bra{n}c_{\alpha}\ket{m}\bra{m}c_{\beta}^\dagger\ket{n}\nonumber\\
\times\frac{e^{-E_n/k_BT} \pm e^{-E_m/k_BT}}{i\nu+E_n-E_m},\nonumber
\end{align}
where the plus (minus) sign is for fermionic (bosonic) frequencies and $Z$ is the canonical partition function. We avoid evaluating at $i\omega = 0$ for the bosonic correlation functions to avoid the singularity.

A useful representation of the above Green's function is given using ``Q-Matricies"\cite{knap2010a}:
$$ Q_{\alpha nm} = \bra{n}c_a\ket{m}\sqrt{\frac{e^{-E_n/k_BT}+e^{-E_m/k_BT}}{Z}} $$
which have dimension ($L^d \times N_{\text{states}}$). We can combine the sum over eigenstates and eigenvalues $n,m$ into a single sum over $s$,
$$G_{\alpha\beta}(i\nu) = \sum_s Q_{\alpha s}\frac{1}{i\nu-\lambda_s}S_{ss}Q_{s\beta}^{\dagger},$$
where $\lambda_s = E_n-E_m$ and the diagonal matrix $S$ handles the statistical sign. We order the Green's function such that hole excitations are first followed by the particle excitations and then
$S=\text{diag}(1,\ldots1,-\xi,\ldots-\xi)$, where $\xi = \mp 1$ is the statistical sign for fermions and bosons, respectively.  This allows a more compact representation:

\begin{align}
\bm{G}(i\nu) = \bm{Q}\frac{1}{i\nu\mathbb{1}-{\bf\Lambda}}\bm{S}\bm{Q}^{\dagger}.
\label{eq_QgQ}
\end{align}

We can now represent the Green's function explicitly in terms of elements of the $Q$ matrices.  Using a Kronecker product we get:
\begin{align}
G_{\alpha\beta}(i\nu) &= \sum_s \left [(\bm{S}\bm{Q}^{\dagger})^{\text{T}}\otimes \bm{Q} \text{vec}\left ( \frac{1}{i\nu-{\bf\Lambda}} \right ) \right ]_s \nonumber\\
&= \sum_s C_{\alpha\beta, s}\frac{1}{i\nu-\lambda_s},
\label{eq_QxQg}
\end{align}
where the subscript $\text{vec}$ indicates transformation of a matrix into a column vector by stacking the 0th column on top of the 1st column, 1st column on top of the 2nd, and so on.  Here we also introduce the rank-3 tensor:

$$ C_{\alpha\beta, s} \equiv \left [(\bm{S}\bm{Q}^{\dagger})^{\text{T}}\otimes \bm{Q} \right]_{\alpha\beta,s}. $$
This expression shows that we can rewrite the Green's function in terms of $Q$ matrices. 

We can also apply the above procedure to the spin susceptibility.  In the Lehman representation we have:

\begin{align}
\chi_{ab}(i\omega) = \frac{1}{Z}\sum_{n,m}\bra{n}S_{a}^+\ket{m}\bra{m}S_{b}^-\ket{n}\nonumber\\
\times\frac{e^{-E_n/k_BT} \pm e^{-E_m/k_BT}}{i\omega+E_n-E_m},\nonumber
\end{align}
which is analogous to the Lehman representation for the Green's function.  We can therefore obtain an expression for the spin susceptibility in terms of $Q$ matrices by replacing fermion operators with spin operators in $ C_{\alpha\beta, s}$.  

The expressions for $G$ and $\chi$ in terms of $Q$-Matrices are useful for efficient evaluation.   For small enough systems we can diagonalize the entire matrix and get all of the wavefunctions, $\ket{\psi_n}$ and $E_n$.  For larger systems we use Lanczos to estimate the Green's function. Using band Lanczos we obtain the Q matrices and the energies $E_n$. At zero temperature, running Lanczos once produces the ground state eigenpairs $\ket{\psi_0}$ and $E_0$.  We then run a banded Lanczos with the set of starting vectors. 
The algorithm produces the $Q$ matrix and the energies $E_m$ allowing approximations of the Green's function and spin susceptibilities. The advantage of this algorithm over normal Lanczos is that there is only a single set of poles for each matrix element.

\section{Lehmann Representation for CPT Green's function}
\label{app_CPT}

We can define the inter-cluster hopping matrix in the mixed representation as:
$$\bm{V}(\tilde{\bm{k}}) = \bm{G}_{0,\text{c}}^{-1}(i\omega) -\bm{G}_{0}^{-1}(\tilde{\bm{k}},i\omega).$$
\noindent
We can see this by splitting up the lattice hopping matrix $\bm{t}$ into an intra and inter-cluster hopping matrix,
$$ \bm{t}(\tilde{\bm{k}}) = \bm{t}_{\text{c}} + \bm{V}(\tilde{\bm{k}}),$$
\noindent
and expressing the non-interacting Green's function for the lattice and the cluster respectively as:
$$ \bm{G}_0^{-1}(\tilde{\bm{k}},i\nu) = i\nu -\bm{t}(\tilde{\bm{k}}), $$
and
$$ \bm{G}_{0,\text{c}}^{-1}(i\nu) = i\nu -\bm{t}_{\text{c}}.$$
This allows us to write the Green's function from the CPT Dyson equation as:
$$\bm{G}_{\text{CPT}}^{-1}(\tilde{\bm{k}},i\nu) = \bm{G}_{\text{c}}^{-1}(i\nu) - \bm{V}(\tilde{\bm{k}}) $$
This is the usual CPT equation as derived in  Ref.~\onlinecite{Senechal2000}. 

\begin{figure}
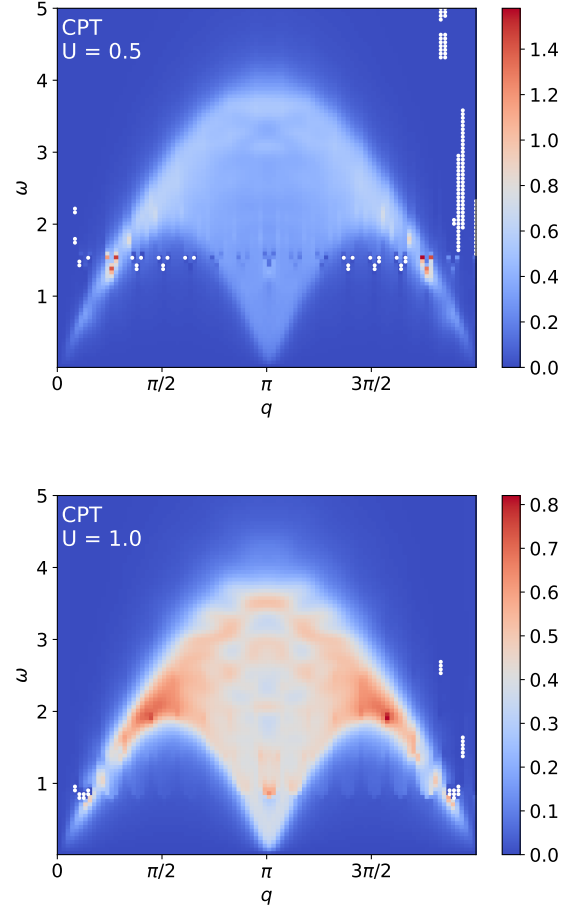

\centering
	\subfloat{\includegraphics[scale=.5]{{{Figures/imchiGL16U0.5P96eps0.2}}}}\\
	\subfloat{\includegraphics[scale=.5]{{{Figures/imchiGL16U1.0P96eps0.2}}}}\\
	\caption{Top: CPT susceptibility for  $L=16$, $U=0.5$ with $\eta = 0.2$ to highlight the numerical instabilities arising from a low broadening parameter. The white dots appear where $\omega < 0$. The badly behaved points follow along an oscillating structure with peaks proportional to cluster size $L$. Bottom: the same but for $U=1.0$. The oscillating structure shrinks as the misbehaving poles lose weight at lower energy scales.  }
\label{fig_badchi}
\end{figure}

If we have the cluster Green's function written in the $Q$-matrix formalism in Sec.~\ref{app_twopoint} we can derive a Lehmann's representation of the CPT Green's function. Following Knap et al.\cite{knap2010a} but ignoring the spin-statistic matrix $S$, since we are only interested in fermions for the Green's function, we get the Lehman representation for the CPT Green's function,

\begin{align}
\bm{G}(\tilde{\bm{k}},i\nu) = \tilde{\bm{Q}}(\tilde{\bm{k}})\frac{1}{i\nu - \tilde{\bf\Lambda}(\tilde{\bm{k}})}\tilde{\bm{Q}}^{\dagger}(\tilde{\bm{k}}),
\label{eq_GCPT_LR}
\end{align}
where the new CPT poles and weights are, respectively:
$$\tilde{\bf\Lambda}(\tilde{\bm{k}}) = \bm{U}(\tilde{\bm{k}})[{\bf\Lambda} +\bm{Q}^{\dagger}\bm{V}(\tilde{\bm{k}})\bm{Q}(\tilde{\bm{k}})]\bm{U}^{\dagger}(\tilde{\bm{k}}) $$
$$\tilde{\bm{Q}}(\tilde{\bm{k}}) = \bm{Q}\bm{U}(\tilde{\bm{k}}).$$
Equation~\ref{eq_GCPT_LR} is our Lehmann representation for the CPT Green's function.

\section{Evaluating the CPT bubble diagram in the mixed representation}
\label{app_bubble}

In this section we rewrite the bubble diagram for the transverse spin susceptibility, $\bm{\chi}_0$, in terms of $Q$-Matrices.  We do this by re-expressing $\bm{\chi}_0$ as a convolution of two Green's functions and analytically computing the frequency integrals.  This provides a more accurate $\bm{\chi}_0$.

In the mixed representation $G_{ab}(\bm{k},i\nu)$ we can write the spin susceptibility as:
$$\chi_{0,ab}({{\bf{q}},i\omega}) = -k_BT\sum_{{\bf{p}},i\nu} G_{ab}({\bf{p}},i\nu)G_{ba}({\bf{p}}+{\bf{q}},i\nu+i\omega).$$
After substituting in the spectral representation:
$$ G_{ab}(\bm{k},i\nu) = \int_{-\infty}^{\infty}d\nu'\frac{-\operatorname{Im}G_{ab}(\bm{k},\nu')}{i\nu-\nu'}, $$
and doing the Matsubara summation,
$$ k_BT\sum_{i\nu}\frac{1}{(i\nu-\nu_1)(i\nu-\nu_2)} = -\xi\frac{f_\xi(\nu_1)-f_\xi(\nu_2)}{\nu_1-\nu_2},$$
where $\xi = \mp 1$ is the statistical sign for fermions and bosons, respectively, and $f_\xi(\nu) = (e^{\nu/k_BT}-\xi)^{-1}$ is either the Fermi-Dirac or Bose-Einstein distribution, we obtain:
\begin{align}
 \chi_{0,ab}({{\bf{q}},i\omega}) &= \nonumber  -\sum_{{\bf{p}}}\iint_{-\infty}^{\infty}d\nu_1d\nu_2 \operatorname{Im}G_{ab}({\bf{p}},\nu_2) \nonumber \\
 &\times\operatorname{Im}G_{ba}({\bf{p}}+{\bf{q}},\nu_2)\frac{f_\xi(\nu_1)-f_\xi(\nu_2)}{i\omega+\nu_1-\nu_2}\nonumber
\end{align}
Passing to the Lehman representation for the Green's function: 
$$ -\operatorname{Im}G_{ab}(\bm{k},\nu) = \sum_{s}C_{ab,s}(\bm{k})\delta(\nu-\lambda_s(\bm{k})), $$
we insert a the Dirac-delta function and do the integral analytically to get:
\begin{align}
\chi_{0,ab}({{\bf{q}},i\omega}) &= -\sum_{{\bm{p}},s,s'}C_{ab,s}(\bm{p})C_{ba,s'}(\bm{p+q})\nonumber \\
&\times \frac{f_\xi(\lambda_s(\bm{p}))-f_\xi(\lambda_{s'}'(\bm{p}+\bm{q}))}{i\omega+\lambda_s(\bm{p})-\lambda_{s'}'(\bm{p}+\bm{q})}.
\end{align}
This is the bubble diagram in mixed representation.  Here $C_{ba,s}$ is defined in terms of $Q$-Matrices in Sec.~\ref{app_twopoint}.

\begin{figure*}
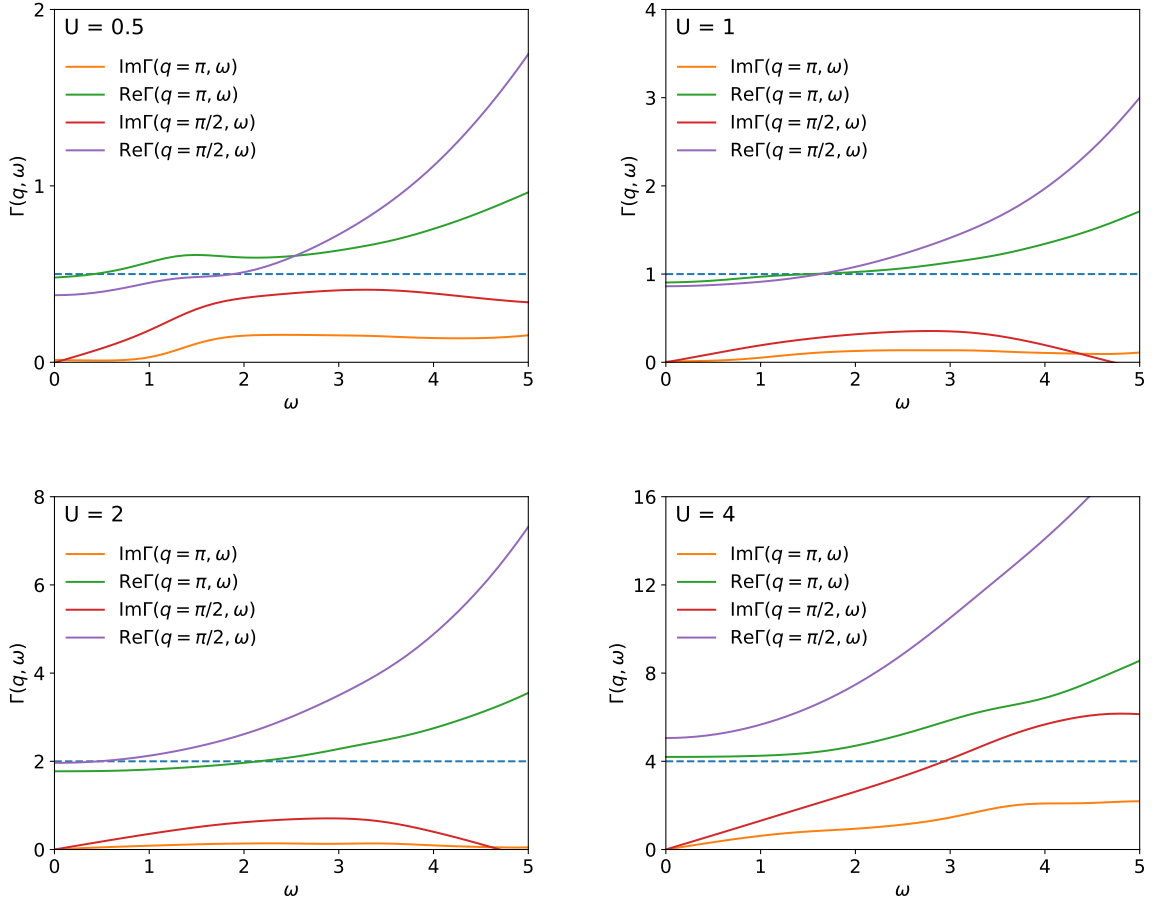

\centering
	\subfloat{\includegraphics[scale=.5]{{{Figures/Vertex/vertex_0p5}}}}
  \subfloat{\includegraphics[scale=.5]{{{Figures/Vertex/vertex_1}}}}\\ 
	\subfloat{\includegraphics[scale=.5]{{{Figures/Vertex/vertex_2}}}}
	\subfloat{\includegraphics[scale=.5]{{{Figures/Vertex/vertex_4}}}}
  \caption{Frequency dependence of the CPT two-point vertex function $\Gamma(q,\omega)$ used in the Bethe-Salpeter equations (See Fig.~\ref{fig_Diagrams}d) for $ H_{\text{1D}}$.  The horizontal dashed lines show the RPA approximation $\Gamma(q,\omega)=U$. The solid lines show the CPT approximation obtained from inverting Eqs.~\ref{eq_twoptBSE} with $\eta = 0.5$ and $N_p = 48$.  Here we see that at low $U$, the real part of the CPT approximation at low frequencies is very close to the RPA.   The imaginary part of the CPT vertex function is also shown for comparison.  }
\label{fig_vertex}
\end{figure*}

\section{Calculation of the CPT Susceptibility}
\label{app_chi}

The momentum resolved spin susceptibility, Eq.~\ref{eq_chiq}, is constructed with imaginary Matsubara frequencies; to compare with experiment, we analytically continue the imaginary frequencies to real frequencies. In general, this analytic continuation is numerically ill-posed. We could apply one of the many analytic continuation algorithms, e.g. Padé approximation\cite{Beach2000}, to Eq.~\ref{eq_chiq}, instead, we chose to analytically continue each of the terms in the CPT spin susceptibility, Eq.~\ref{eq_chiqCPT} separately. Before the inversion, each term is the sum of simple poles. Therefore, the analytic continuation is straightforward and given by replacing $i\omega \rightarrow \omega +i\eta$, where $\eta>0$ is some small parameter.

In general none of the terms in Eq.~\ref{eq_chiqCPT} have to be invertible. In particular we find that $\bm{\chi}_\text{c}$ has a zero eigenvalue for all $\omega$'s and $U$'s due to being in the paramagnetic phase and not breaking the SU(2) symmetry\cite{Filor2014a}. One way to proceed is to regularize the ill-conditioned matrix inversion by adding a small parameter along the diagonal. Another way is to twice apply the matrix identity:
$$(\bm{A}^{-1}+\bm{B}^{-1})^{-1} = \bm{A}(\bm{A}+\bm{B})^{-1}\bm{B},$$
where $\bm{A},\bm{B},$ and $\bm{C}$ are matrices and $(\bm{A}+\bm{B})$ is assumed to be invertible. These two methods agree provided that the regularization parameter is sufficiently small. This avoids the explicit calculation of the vertex $\bm{\Gamma}_c$ and instead calculates $\bm{\Gamma}_c^{-1}$.

Unfortunately, this method did not produce the correct behavior at low $U$ when compared with RPA-CPT. We instead attempt to find the best $\bm{\Gamma}_c$ such that Eq.~\ref{eq_twoptBSEc} is satisfied. This is accomplished via pseudoinverses and is given by:
\begin{align}
\bm{\Gamma}_c = \bm{\chi}_{0,c}^{+}(\bm{\chi}_c-\bm{\chi}_{0,c})\bm{\chi}_c^{+}.
\end{align}
Using the pseudoinverse matrices is equivalent to finding the least squares solution. The pseudoinverse can be found with a singular value decomposition,
$$ \bm{A} = \bm{U}\bm{\Sigma}\bm{V}^{\dagger} $$
for any $(\bm{A}_{\text{rows}} \times \bm{A}_{\text{cols}})$ matrix $\bm{A}$. Where $\bm{U}$ and $\bm{V}$ are unitary matrices with sizes $(\bm{A}_{\text{rows}} \times \bm{A}_{\text{rows}})$ and $(\bm{A}_{\text{cols}} \times \bm{A}_{\text{cols}})$ respectively. $\bm{\Sigma}$ is a diagonal $(\bm{A}_{\text{rows}} \times \bm{A}_{\text{cols}})$ matrix with non-negative real numbers. The pseudoinverse is given by:
$$\bm{A}^{+} = \bm{V}\bm{\Sigma}^{+}\bm{U}^{\dagger},$$
where $\bm{\Sigma}^{+}$ is found by replacing the non-zero elements with their reciprocals and transposing the resultant matrix. This effectively ignores the problematic zero eigenvalue that occurs in $\bm{\chi}_{c}$.

\section{Causality and Bounds on $\eta$}
\label{app_numerical}

The CPT applied to the spin susceptibility as written in Eq.\ref{eq_chiqCPT} does not strictly respect causality ($\chi(\bm{q},\omega) > 0 $ for $\omega > 0$ and $\chi(\bm{q},\omega) < 0 $ for $\omega < 0$). This is due to the minus sign on $\bm{\chi}_{0,\text{c}}(\omega)^{-1}$. While we can derive a Lehman representation for each term in Eq.~\ref{eq_chiqCPT}, the poles are derived independently and therefore differ in finite sized systems (the poles match in the thermodynamic limit). Slight differences in the poles lead to unphysical divergences, causality violations, and periodic numerical artifacts in the CPT susceptibility as shown in Fig.~\ref{fig_badchi}. 

By picking a large enough broadening parameter $\eta$ we can minimize these issues. The value of $\eta$ needed depends on the interaction strength $U$. For $U<2$, this numerical issue is the most severe except for at $U=0$ where $\bm{\chi}_{0,\text{c}}(\omega) = \bm{\chi}_{\text{c}}(\omega)$ exactly. For larger interaction strength $U>4$, $\bm{\chi}_\text{CPT}$ is dominated by $\bm{\chi}_{0,\text{c}}(\omega)$ and the other terms are pushed off to higher energies.   We understand this by noting that around $U\sim 0$, we have $\bm{\chi}_c \sim \bm{\chi}_{0,c} \sim \bm{\chi}_0$, and therefore, the slight differences between the many poles cause numerical artifacts. A larger $\eta$ smooths out the spectra and alleviates the unphysical divergences and signs but still retains some of the numerical artifacts.

Numerical artifacts in  CPT applied to two-particle correlation functions can be removed entirely with methods that are beyond the scope of the present work. One approach would be to derive a Lehmann representation of $\bm{\chi}_\text{CPT}$, like the one for the Green's function (Eq.~\ref{eq_GCPT_LR}) , then to combine poles until causality is respected\cite{Lu2014}.

\section{Vertex Function Comparison}
\label{app_vertex}

This section compares the two-point vertex function $\Gamma(\bm{q},i\omega)$ computed with CPT against the RPA approximation for a few values of $\bm{q}$.  We first note that only frequencies below a certain threshold are relevant, i.e., only $\omega < 4$ is important in Figs.~\ref{fig_lowU}-\ref{fig_U4}, since the susceptibility vanishes for  $\omega > 4$.   We are therefore interested in comparing vertex functions only for $\omega < 4$.

We calculate $\Gamma(\bm{q},i\omega)$ by first computing $\chi(\bm{q},i\omega)$ via Eq.~\ref{eq_chiq} and $\chi_0(\bm{q},i\omega)$ with the Green's function from Eq.~\ref{eq_G_CPT_q}. By inverting the two-point Bethe-Salpeter equation in momentum space,
$$\chi(\bm{q},i\omega) = \chi_0(\bm{q},i\omega) + \chi_0(\bm{q},i\omega)\Gamma(\bm{q},i\omega)\chi(\bm{q},i\omega),$$
we can construct a momentum resolved vertex $\Gamma(\bm{q},i\omega)$.

\begin{figure}[t]
\vspace{1cm}
\centering
	\subfloat{
\includegraphics[scale=0.45
,angle=0]{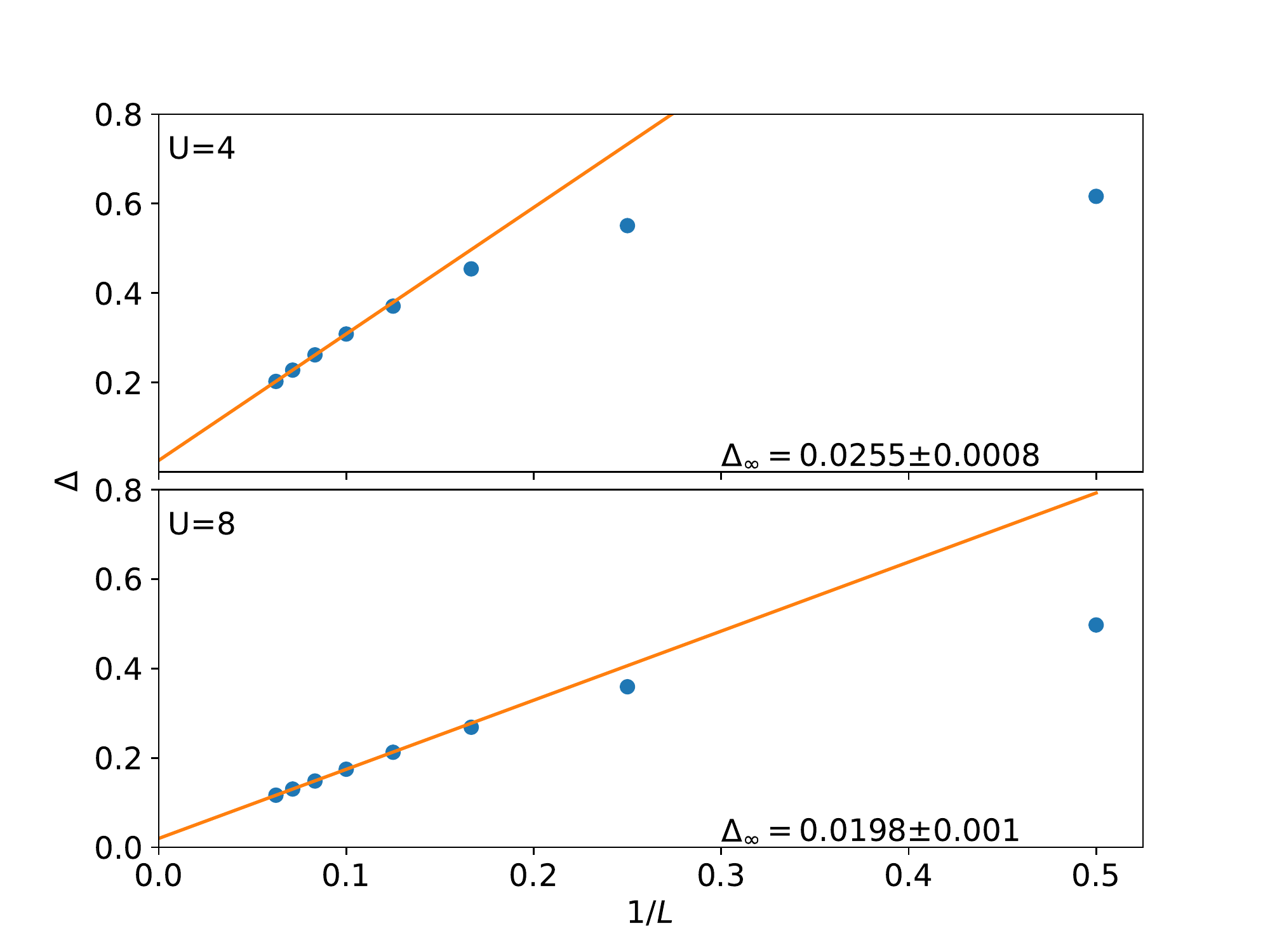}}
	\caption{Finite-size trend of the gap $\Delta$.  $\Delta$ is obtained as the frequency for which $\chi(\pi,\omega)$ is first non-zero using CPT for $U=4$ (top) and $U=8$ (bottom) in $H_{\text{1D}}$.  Here we see a trend toward zero.  Linear fits of the last four data points show $y$-intercepts of $0.0255(8)$ and $0.020(1)$ for $U=4$ and $U=8$, respectively.   }
\label{fig_scaling}
\end{figure}

The solid lines in Fig.~\ref{fig_vertex} plot the imaginary and real parts of the vertex function versus frequency for several values of $U$. The dashed lines plot the RPA approximation which should compare well with CPT at low $U$.  The RPA approximation to the vertex function is real and independent of $q$ and $\omega$.

From Fig.~\ref{fig_vertex} we see that for $U=1$ and $U=2$ the CPT approximation shows reasonable agreement with RPA for the relevant frequency ranges, $\omega < 3$ and $\omega < 4$, respectively.  The $U=4$ panel shows that CPT and RPA start to deviate significantly as expected since RPA is valid in the low $U$ limit.  Fig.~\ref{fig_vertex} also shows that for $U=0.5$ we have a peak in the vertex function.  This peak is entirely numerical in origin and derives from the pole mismatch discussed in Secs.~\ref{app_chi} and \ref{app_numerical}.

\section{Finite-Size Scaling of the $\chi(\pi,\omega)$ Gap}
\label{app_scaling_gap}

We define the gap in spin susceptibility $\Delta$ to be  the frequency at which $\chi(\pi,\omega)$ attains a non-zero value.  $\Delta$  corresponds to the energy of the lowest excitation. It should scale to zero in the thermodynamic limit since the spectrum of the 1D Hubbard model is gapless in the thermodynamic limit \cite{Lieb2003}.  Specifically, the $\chi(\pi,\omega\rightarrow 0)$ gap in Fig.~\ref{fig_U4} should go to zero as we increase $L$.  The precise scaling of the energy gap depends on $U$ and the boundary conditions \cite{Lieb2003}.  We focus on larger values of $U$ since the Heisenberg limit is known from analytic arguments to scale as one over the system size \cite{Karbach1997} which motivates an expected $1/L$ scaling in our CPT study.  

Fig.~\ref{fig_scaling} plots the energy gap $\Delta$ as function of $1/L$ to show a clear diminishing trend.  But we find that a linear fit to the 4 data points at the largest $L$ yield a small but non-zero gap: $\Delta_{\infty}=0.0255(8)$ for $U=4$ and $\Delta_{\infty}=0.020(1)$ for $U=8$.  The non-zero extrapolations may be due to the small system size used, $L\leq 16$.  We note that in other studies of related two-spin correlation functions, DMRG \cite{Hallberg1995,NISHIMOTO2007} was needed to extract the correct finite size scaling from larger system sizes (more than 70 sites) in the Heisenberg model.  We conclude that while the  $\chi(\pi,0)$ gap diminishes, our system sizes are too small to extract a zero gap in finite-size scaling.

\section{Estimating the Chemical Potential Away from Half Filling}
\label{app_chemical_potential}

We estimate the chemical potential in the 1D Hubbard model by varying $\mu$ until the density $n$ reaches the desired value.  Specifically, we start with the CPT Green's function, Eq.~\ref{eq_G_CPT_q}, and use this to get the density of states:
\begin{align}
\rho(\nu) = \frac{-1}{N_p} \sum_{\bf{k}} [\operatorname{Im}G({\bf{k}},\nu)].
\label{dos}
\end{align}
By noting that we must have $\int_{-\infty}^{\infty} d\nu \rho(\nu) = 1$, we are able to find the correct chemical potential, $\mu$, for the desired density via, 
\begin{align}
n = \int_{-\infty}^{\infty}d\nu f(\nu) \rho(\nu),
\label{particledensity}
\end{align}
where $f(\nu)$ is the usual Fermi-Dirac distribution function.  We approximate the chemical potential, $\mu$, by setting $n=n_c$, the fixed density on the cluster, then using a root finder on equation \ref{particledensity} with $f(\nu) = 1-\Theta(\nu-\mu)$ at $T=0$.

\begin{table}[t]
  \vspace{1cm}
  \centering
  \begin{tabular}{|c|c|c|c|c|c|c|}
    \hline
    $n_c/2$ & \multicolumn{2}{|c|}{1/8} & \multicolumn{2}{c|}{1/4} & \multicolumn{2}{c|}{1/2}\\
    \hline
    $U$ & $\mu$ & $n/2$ & $\mu$ & $n/2$ & $\mu$ & $n/2$ \\
    \hline
    0 & -1.850 & 0.128 & -1.421 & 0.248 & 0.019 & 0.500 \\
    1 & -1.730 & 0.127 & -1.205 & 0.251 & 0.461 & 0.499 \\
    2 & -1.680 & 0.128 & -1.025 & 0.251 & 0.828 & 0.499 \\
    4 & -1.589 & 0.127 & -0.739 & 0.249 & 1.257 & 0.500 \\
    8 & -1.531 & 0.128 & -0.491 & 0.251 & 2.744 & 0.500 \\
    \hline
  \end{tabular}
  \caption{Computed chemical potentials, $\mu$, and densities, $n$, for each interaction strength $U$ for one-eighth, one-quarter, and half filling, respectively, using $N_{p} = 500$.  The top row indicates the filling fixed on the cluster.}
  \label{chem_list}
  \vspace{1cm}
\end{table}

After following the above steps to approximate the chemical potential we use the obtained value of $\mu$ in $H_{\text{1D}}$  such that the correct particle density occurs at $\nu = 0$ in our Green's function.  Example chemical potentials are listed in table \ref{chem_list}.

\bibliography{CPT1dHubbard}

\end{document}